\documentclass[a4paper]{article}
\usepackage{graphicx}

\usepackage[english]{babel}
\usepackage[utf8]{inputenc}
\usepackage{geometry}
\usepackage{amsmath}
\usepackage{amssymb}
\usepackage{epstopdf}
\usepackage{upgreek}
\usepackage{microtype}
\usepackage{color}
\usepackage{booktabs}
\usepackage{authblk}
\usepackage{lineno}

\title{Neutral-current neutrino-nucleus scattering off Xe isotopes}

\author[1]{P. Pirinen}
\author[1]{J. Suhonen}
\author[2]{E. Ydrefors}

\affil[1]{\small University of Jyvaskyla,
Department of Physics, P. O. Box 35 (YFL), FI-40014, Finland \normalsize}
\affil[2]{\small Instituto Tecnol\'ogico de Aeron\'autica, DCTA, 12228-900 S\~ao Jos\'e dos Campos, Brazil \normalsize}
\begin{document}

\maketitle

\begin{abstract}
Large liquid xenon detectors aiming for dark matter direct detection will soon become viable tools also for investigating neutrino physics. Information on the effects of nuclear structure in neutrino-nucleus scattering can be important in distinguishing neutrino backgrounds in such detectors. We perform calculations for differential and total cross sections of neutral-current neutrino scattering off the most abundant xenon isotopes.  The nuclear structure calculations are made in the nuclear shell model for elastic scattering, and also in the quasiparticle random-phase approximation (QRPA) and microscopic quasiparticle phonon model (MQPM) for both elastic and inelastic scattering. Using suitable neutrino energy distributions, we compute estimates of total averaged cross sections for $^{8}$B solar neutrinos and supernova neutrinos.
\end{abstract}

\section{Introduction}

When the idea of neutrinos was first suggested by Pauli in 1930, it was thought that they would never be observed experimentally. Only two decades later interaction of neutrinos with matter was detected in the famous Cowan-Reines experiment \cite{Cowan1956}. More recently, detection and research of neutrinos has become more and more of an everyday commodity, and various more versatile ways to examine interactions of the little neutral one have emerged and are being tested in laboratories all over the world.

Coherent elastic neutrino-nucleus scattering (CE$\nu$NS) is a process where the neutrino interacts with the target nucleus as a whole instead of a single nucleon. Although CE$\nu$NS has been predicted since the 1970s \cite{Freedman1974}, it was discovered only very recently by the COHERENT collaboration \cite{Akimov2017}. Due to the coherent enhancement, this experiment had the remarkable feature of detecting neutrinos with a compact 14.6 kg detector instead of a massive detector volume which is used in conventional neutrino experiments. Coherent neutrino nucleus scattering is on one hand an important potential source of information for beyond-standard-model physics  \cite{Anderson2012, Dutta2016a, Kosmas2017,Kosmas2015,Barranco2007, deNiverville2015, Dutta2016b,Lindner2017}, but on the other hand it may also hinder new discoveries as it will start disturbing dark matter detectors in the near future.

A great experimental effort has been put into directly detecting dark matter in the past few decades  (see Ref. \cite{Undagoitia2016} for a review). The next-generation detectors are expected to be sensitive enough to probe cross sections low enough to start observing CE$\nu$NS as an irreducible background  \cite{Monroe2007,Billard2014}. Solar neutrinos, atmospheric neutrinos, and diffuse supernova background neutrinos provide a natural source of background neutrinos, which for obvious reasons cannot be shielded against. As there are uncertainties in the fluxes of each of the aforementioned neutrino types, the sensitivity of WIMP detection is basically limited to the magnitude of this uncertainty. To make matters worse, it has been shown that for some specific WIMP masses and cross sections the recoil spectra of CE$\nu$NS very closely mimics that of scattering WIMPs \cite{Billard2014}.   

It is therefore of utmost importance to device a way to go through this neutrino floor. One potential way of achieving this is having directional sensitivity in the detector \cite{OHare2015,Grothaus2014}. As solar and atmospheric neutrinos have a distinct source within the solar system, it is expected that their recoil direction would be different of that of WIMPs, which are typically assumed to be gravitationally bound in a halo spanning the galaxy. Also arising from the different origin of neutrinos and WIMPs is the idea of using timing information to discriminate between neutrino and WIMP induced events in a detector \cite{Davis2015}. Due to the motion of the Earth around the Sun, it is expected that the solar neutrino flux peaks around January, but the WIMP flux peaks in June when the velocities of the Sun and Earth are the most in phase. The recoil spectra of WIMPs and neutrinos could also be distinquished if the WIMP-nucleus interaction happens via a nonstandard operator emerging in the effective field theory framework \cite{Dent2016, Dent2017}.

Some of the leading dark matter experiments use a liquid xenon target  \cite{XENON2017,XMASS2013,LZ2015,LUX2017,DARWIN2016}, which allows for easy scalability to larger detector volumes. It is expected that the xenon detectors are the first to hit the neutrino floor. In this article we compute cross sections for elastic and inelastic neutrino-nucleus scattering for the most abundant xenon isotopes. For the coherent scattering we use the quasiparticle random-phase approximation (QRPA) framework and the nuclear shell model to model the nuclear structure and we compare the results between the two models. The wave functions of the states of odd-mass xenon isotopes are obtained by using the microscopic quasiparticle-phonon model (MQPM) on top of a QRPA calculation. Inelastic scattering is computed in the QRPA/MQPM formalism. In our calculations we consider $^{8}$B solar neutrinos and supernova neutrinos.  

A similar QRPA calculation has been made in Ref. \cite{Ydrefors2015} for $^{136}$Xe, where both charged-current and neutral-current inelastic scattering was examined. Similar computations of neutral-current neutrino-nucleus scattering cross sections have been made before for the stable cadmium isotopes in Ref. \cite{Almosly2015}, and for molybdenum isotopes in Ref. \cite{Ydrefors2012}. Both calculations used the QRPA/MQPM approach. To our knowledge this article presents the first calculation of neutral-current neutrino-nucleus scattering within a complete microscopic nuclear framework for Xe isotopes other than $^{136}$Xe.

This article is organized as follows. In section \ref{sec:ncscatt} we outline the formalism used to compute neutral-current neutrino-nucleus scattering. In section \ref{sec:nucl} we summarize the nuclear-structure calculations made for the target xenon isotopes. In section \ref{sec:results} we discuss the results of our cross-section calculations and in section \ref{sec:conc} conclusions are drawn. 

\section{Neutral-current neutrino-nucleus scattering} \label{sec:ncscatt}

\begin{figure}
\center
\includegraphics[width=0.5\textwidth]{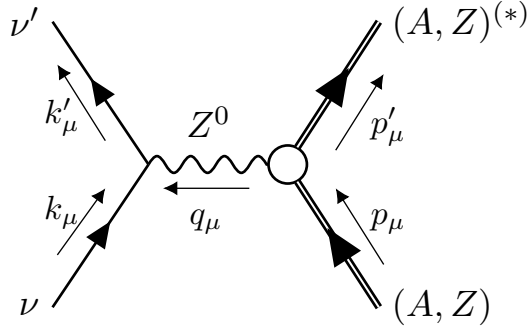}
\caption{A diagram of the neutral-current scattering process. The four-momenta of the involved particles are labeled in the figure.  } \label{fig:kinematics}
\end{figure}

In this section we summarise the formalism used to compute neutral-current neutrino-nucleus scattering processes. We examine standard-model reactions mediated by the neutral $Z^0$ boson, namely the processes 
\begin{align}
\nu & + (A,Z) \rightarrow \nu + (A, Z), \\
\nu & + (A,Z) \rightarrow \nu + (A, Z)^* ,
\end{align}
i.e. the elastic and inelastic scattering of neutrinos off a nucleus (with $A$ nucleons and $Z$ protons), respectively. In the elastic process the initial and final states of the target nucleus are the same, while in the inelastic process excitation of the target nucleus takes place. The kinematics of the scattering process is illustrated in Fig. \ref{fig:kinematics}. We label the four-momenta of the incoming and outgoing neutrino as $k_\mu$ and $k_\mu^\prime$. The momenta of the target nucleus before and after interacting with the neutrino are $p_\mu$ and $p_\mu^\prime$. The momentum transfer to the nucleus is referred to as $q_\mu = k_\mu^\prime-k_\mu = p_\mu - p_\mu^\prime $. The neutrino kinetic energy before and after scattering is $E_k$ and $E_{k^\prime}$

The neutral-current neutrino-nucleus scattering differential cross section to an excited state of energy $E_\mathrm{ex}$ can be written as \cite{Ydrefors2011}
\begin{equation}
\frac{d^2 \sigma}{d\Omega dE_\mathrm{ex}} = \frac{G_\mathrm{F}^2 \left| \mathbf{k^\prime} \right| E_{k^\prime}}{\pi (2J_i+1)}\left( \sum_{J\geq 0} \sigma_\mathrm{CL}^J + \sum_{J\geq 1} \sigma_\mathrm{T}^J \right),
\end{equation}
which comprises of the Coulomb-longitudinal $(\sigma_\mathrm{CL}^J)$ and transverse $(\sigma_\mathrm{T}^J)$ parts. They are defined as
\begin{align*}
\sigma_\mathrm{CL}^J =  &(1+\cos \theta)\left| \langle J_f||\mathcal{M}_J(q)||J_i \rangle  \right|^2+(1+\cos \theta - 2b\sin^2 \theta)\left| \langle J_f || \mathcal{L}_J(q) || J_i \rangle \right|^2  \\
 & + qE_\mathrm{exc}(1+\cos \theta) \times 2 \mathrm{Re}\left\{ \langle J_f || \mathcal{M}_J(q) || J_i \rangle^*  \langle  J_f || \mathcal{L}_J(q) || J_i \rangle \right\},
\end{align*}
and 
\begin{align*}
\sigma_\mathrm{T}^J =& \left( 1- \cos \theta + b \sin ^2 \theta \right) \left[ \left| \langle J_f || \mathcal{T}_J^\mathrm{mag}(q) || J_i \rangle \right|^2 + \left| \langle J_f || \mathcal{T}_J^\mathrm{el}(q) \rangle \right|^2 \right] \\ & \mp \frac{E_k + E_{k^\prime}}{q} (1- \cos \theta)\times 2 \mathrm{Re} \left\{ \langle J_f || \mathcal{T}_J^\mathrm{mag}(q) || J_i \rangle  \langle J_f || \mathcal{T}_J^\mathrm{el} || J_i  \rangle \right\},
\end{align*}
where the minus sign is taken for antineutrino scattering and the plus sign for neutrino scattering. $J_i$ and $J_f$ are the initial and final state angular momenta of the nucleus. We use the abbreviation 
\begin{equation}
b = \frac{E_\mathbf{k} E_\mathbf{k^\prime}}{q^2},
\end{equation}
and $q$ is the magnitude of the three-momentum transfer. The formalism and various different operators involved are discussed in detail in Refs. \cite{Ydrefors2011,Donnelly1979}.

To compute the averaged cross section $\langle \sigma \rangle$, we need to fold the computed cross sections with the energy distribution of the incoming neutrinos. We take the supernova neutrino spectrum to be of a two-parameter Fermi-Dirac character
\begin{equation}\label{eq:fd}
f_\mathrm{FD}(E_k) = \frac{1}{F_2(\alpha_\nu)T_\nu} \frac{(E_k/T_\nu)^2}{1+e^{E_k/(T_\nu-\alpha_\nu)}},
\end{equation}
where $\alpha_\nu$ is the so-called pinching parameter, and $T_\nu$ the neutrino temperature. The normalization factor $F_2(\alpha_\nu)$ is defined by the formula 
\begin{equation} \label{eq:ffactor}
F_k(\alpha_\nu) = \int \frac{x^k dx}{1+ e^{x-\alpha_\nu}},
\end{equation}
and the temperature and mean energy of neutrinos are related by
\begin{equation} \label{eq:temperel}
\frac{\langle E_\nu \rangle}{T_\nu} = \frac{F_3(\alpha_\nu)}{F_2(\alpha_\nu)}.
\end{equation}
We also examine solar neutrinos from $^{8}$B beta decay. We use an $^{8}$B neutrino energy spectrum from Ref. \cite{Bahcall1996}.

\section{Nuclear structure of the target nuclei}\label{sec:nucl}

In this section we outline the nuclear-structure calculations performed for the investigated nuclei $^{128,129,130,131,132,134,136}$Xe. We have performed computations in the quasiparticle random-phase approximation (QRPA), microscopic quasiparticle-phonon model (MQPM), and the nuclear shell model. 

\subsection{QRPA/MQPM calculations}

The nuclear structure of even--even Xe isotopes was computed by using the charge-conserving QRPA framework. The QRPA is based on a BCS calculation, where quasiparticle creation and annihilation operators are defined via the Bogoliubov-Valatin transformation as
\begin{align*}
a_\alpha^\dag = & u_a c_\alpha^\dag + v_a \tilde{c}_\alpha, \\
\tilde{a}_\alpha = & u_a \tilde{c}_\alpha - v_a c_\alpha^\dag,
\end{align*}
with the regular particle creation and annihilation operators $c_\alpha^\dag$ and $\tilde{c}$ defined in \cite{Suhonen2007}. Here $\alpha$ contains the quantum numbers $(a, m_\alpha)$ with $a = (n_a, l_a, j_a)$. The excited states with respect to the QRPA vacuum are created with the phonon creation operator 
\begin{equation}\label{eq:qrpaphonon}
Q_\omega^\dag = \sum_{ab} \mathcal{N}_{ab}(J_\omega) \left( X_{ab}^\omega [ a_a^\dag a_b^\dag ]_{J_\omega M_\omega} + Y_{ab}^\omega [ \tilde{a}_a \tilde{a}_b]_{J_\omega M_\omega} \right)
\end{equation}
for an excited state $\omega = (J_\omega,M_\omega,\pi_\omega,k_\omega)$, where $k_\omega$ is a number labeling the excited states of given $J^\pi$. In the above equation 
\begin{equation}\label{eq:norm}
\mathcal{N}_{ab}(J_\omega) = \frac{ \sqrt{1+\delta_{ab}(-1)^{J_\omega}}}{1+\delta_{ab}},
\end{equation}
and $X_{ab}^\omega$ and $Y_{ab}^\omega$ are amplitudes describing the wave function that are solved from the QRPA equation
\begin{equation}
\begin{bmatrix}
\mathrm{A} & \mathrm{B} \\
-\mathrm{B}^* & -\mathrm{A}^*
\end{bmatrix}
\begin{bmatrix}
\mathrm{X}^\omega \\
\mathrm{Y}^\omega
\end{bmatrix}
= E_\omega
\begin{bmatrix}
\mathrm{X}^\omega \\
\mathrm{Y}^\omega
\end{bmatrix},
\end{equation}
where the matrix A is the basic Tamm-Dankoff matrix and B is the so called correlation matrix, both defined in detail in \cite{Suhonen2007}. 

We perform the QRPA calculations using large model spaces consisting of the entire 0$s$--0$d$, 1$p$--0$f$--0$g$, 2$s$--1$d$--0$h$, and 1$f$--2$p$ major shells, adding also the $0i_{13/2}$ and $0i_{11/2}$ orbitals. The single-particle bases are constructed by solving the Schrödinger equation for a Coulomb-corrected Woods-Saxon potential. We use the Woods-Saxon parameters given in Ref. \cite{Bohr1969}. We make an exception for $^{136}$Xe, adopting the set of adjusted values of single-particle energies from Ref. \cite{Ydrefors2015}. Due to the neutron-magic nature of $^{136}$Xe, adjusted single-particle energies are necessary to get agreement with experimental energy levels. The Bonn one-boson exchange potential \cite{Holinde1981} was used to estimate the residual two-body interaction. 

The QRPA formalism involves several parameters that have to be fixed by fitting observables to experimental data. In the BCS calculation we fit the proton and neutron pairing strengths $A_\mathrm{pair}^\mathrm{p}$ and $A_\mathrm{pair}^\mathrm{n}$ so that the lowest quasiparticle energy matches the empirical pairing gap given by the three-point formula \cite{Wapstra1985}:
\begin{align*}
\Delta_{\rm{p}}(A, Z) &= \frac{1}{4} (-1)^{Z+1}\big[ S_{\rm{p}}(A +1, Z +1 ) - 2S_{\rm{p}}(A, Z) + S_{\rm{p}}(A-1, Z-1)\big],  \\
\Delta_{\rm{n}}(A, Z) &= \frac{1}{4} (-1)^{A - Z+1}\big[ S_{\rm{n}}(A +1, Z ) - 2S_{\rm{n}}(A, Z) + S_{\rm{n}}(A-1, Z)\big].
\end{align*}
It should be noted, that for the neutron-magic $^{136}$Xe this procedure cannot be done for the neutron pairing strength. We have instead used a bare value of  $A_\mathrm{pair}^\mathrm{n} = 1.0$ for $^{136}$Xe. 

The particle--particle and particle--hole terms of the two-body matrix elements are scaled by strength parameters $G_\mathrm{pp}$ and $G_\mathrm{ph}$, respectively. The energies of the computed QRPA states are quite sensitive to these model parameters. We fit the lowest excited states of each $J^\pi$ separately to experimental values from Ref. \cite{nndc} by altering the values of $G_\mathrm{pp}$ and $G_\mathrm{ph}$. The values used for the model parameters are given in Table \ref{tab:mparam}.

\begin{table}[htbp]
\centering
\caption{Model parameters used in the BCS and QRPA calculations. For each nucleus (column 1) the values of $G_\mathrm{pp}$ and $G_\mathrm{ph} $ (column 2) are given for the important $J^\pi$ phonons in columns 3 to 9.} 
\label{tab:mparam}
\begin{tabular}{lllllllll} 
	\toprule \toprule

	 Nucleus &  $G $    & $0^+$  & $1^-$ & $2^+$ & $3^-$ & $4^+$ & $5^-$  & $6^+$ \\
	\midrule
           $^{128}$Xe & $\mathrm{pp}$ & 0.796 & 1.000 & 1.000 & 1.000 & 1.000 & 1.000  & 1.000 \\
			 & $\mathrm{ph} $ & 0.298 & 0.500 & 0.527 & 0.500 & 0.652 & 0.883 & 0.934 \\
           $^{130}$Xe & $\mathrm{pp}$ &  0.730 & 1.000 & 1.000 & 1.000 & 1.000 & 1.000 & 1.000 \\
			 & $\mathrm{ph} $ &  0.303 & 0.500 & 0.531 & 0.500 & 0.581 & 0.833 & 0.788\\
           $^{132}$Xe & $\mathrm{pp}$ &  0.653 & 1.000 & 1.000 & 1.000 & 1.000 & 1.000 & 1.000 \\
			 & $\mathrm{ph} $ & 0.319 & 0.500 & 0.533 & 0.500 & 0.436 & 0.933 & 1.000  \\
           $^{134}$Xe & $\mathrm{pp}$ &  0.500 & 1.000 & 1.000 & 1.000 & 1.000 & 1.000 & 1.000 \\
			 & $\mathrm{ph} $ &  0.370 & 0.500 & 0.511 & 0.500 & 0.596 & 1.000 & 0.891 \\
           $^{136}$Xe & $\mathrm{pp}$ &  0.843 & 1.000 & 1.000 & 1.000 & 1.000 & 1.000 & 1.000 \\
			 & $\mathrm{ph} $ &  0.100 & 0.500 & 0.583 & 0.500 & 0.700 & 0.747 & 0.891 \\
     
\bottomrule \bottomrule
\end{tabular}
\end{table}

The QRPA process is known to produce states that are spurious, namely the first excited $0^+$ state and the first $1^-$ state. The first $0^+$ state has been deemed spurious in Refs. \cite{Almosly2015,Almosly2013}. The first $1^-$ state is due to spurious center-of-mass motion as described in Ref. \cite{Suhonen2007}. We have fitted the energies of these states to zero, if possible, by using the model parameters $G_\mathrm{pp}$ and $G_\mathrm{ph}$, and subsequently the states have been omitted from calculations for the even-mass isotopes and also from the MQPM calculations for the odd-mass isotopes. The contributions of these spurious states to the total neutrino-nucleus scattering cross section would be tiny in any case. 

Odd-mass xenon isotopes $^{129,131}$Xe were then computed by using the MQPM formalism, in which we use a combination of one- and three-quasiparticle states by coupling a quasiparticle with a QRPA phonon to form the three-quasiparticle configurations. The MQPM basic excitation can be written in terms of quasiparticle and QRPA-phonon creation operators as \cite{Toivanen1998}
\begin{equation}\label{eq:mqpmex}
\Gamma_k^\dag (jm) = \sum_n C_n^k a_{njm}^\dag + \sum_{a,\omega} D_{a\omega}^k \left[ a_a^\dag Q_\omega^\dag \right]_{jm} .
\end{equation}
The amplitudes $C$ and $D$ are computed by solving the MQPM equations of motion. The detailed description of the process can be found in Ref. \cite{Toivanen1998}. No additional model parameters are required for the MQPM calculation aside for the parameters fitted for the BCS/QRPA calculation described above. We do the MQPM calculations of $^{129}$Xe and $^{131}$Xe using $^{130}$Xe and $^{132}$Xe as reference nuclei, respectively. We select all QRPA phonons of $J \leq 6$ with an energy less than 10 MeV to be used in the calculation.

\subsection{Shell-model calculations}

We perform shell model calculations for Xe isotopes using the shell-model code NuShellX@MSU \cite{Brown2014}. We use the $0g_{7/2}$, $1d_{5/2}$, $1d_{3/2}$, $2s_{1/2}$, and $0h_{11/2}$ valence space and the SN100PN interaction \cite{Brown2005}. The single-particle energies associated with the aforementioned orbitals in the SN100PN interaction are 0.8072, 1.5623, 3.3160, 3.2238, and 3.6051 MeV, respectively, for protons, and $-10.6089$, $-10.2893$, $-8.7167$, $-8.6944$, and $-8.8152$ MeV for neutrons. 

The matrix dimension in the shell-model calculation increases rapidly when moving away from the $N = 82$ shell closure of $^{136}$Xe. For $^{132,134,136}$Xe we were able to do a full calculation with no truncations, but for $^{128-131}$Xe we had to put restrictions on the neutron valence space. The truncations made for each isotope are shown in detail in Table \ref{tab:xetrunc}. For the isotopes $^{128-131}$Xe we assume a completely filled $0g_{7/2}$ orbital and for $^{128,129,131}$Xe we also assume the $1d_{5/2}$ orbital to be full. These should be reasonable approximations when aiming to describe the ground state and low-lying excited states in the xenon nuclei. The orbitals $0g_{7/2}$ and $1d_{5/2}$ have the lowest single-particle energies and the excitations are likely to take place from higher orbitals when the neutron number of the nuclei is quite large. 

\begin{table}[htbp]
\center
\caption{The valence-space truncations made in the shell-model calculations. The first column labels the Xe isotope, the following five columns give the minimum/maximum number of neutrons on the single-particle orbitals $0g_{7/2}$, $1d_{5/2}$, $1d_{3/2}$, $2s_{1/2}$ , and $1h_{11/2}$, respectively. } \label{tab:xetrunc}
\begin{tabular}{llllll}
\toprule \toprule
Nucleus & $0g_{7/2}$ & $1d_{5/2}$ & $1d_{3/2}$ & $2s_{1/2}$ & $1h_{11/2}$ \\
\midrule
$^{128}$Xe  & 8/8 & 6/6 & 0/4 & 0/2 & 4/12 \\
$^{129}$Xe  & 8/8 & 6/6 & 0/4 & 0/2 & 4/12 \\
$^{130}$Xe  & 8/8 & 4/6 & 0/4 & 0/2 & 0/12 \\
$^{131}$Xe  & 8/8 & 6/6 & 0/4 & 0/2 & 0/12 \\
$^{132}$Xe  & 0/8 & 0/8 & 0/4 & 0/2 & 0/12 \\
$^{134}$Xe  & 0/8 & 0/8 & 0/4 & 0/2 & 0/12 \\
$^{136}$Xe  & 0/8 & 0/8 & 0/4 & 0/2 & 0/12 \\
\bottomrule \bottomrule

\end{tabular}
\end{table} 

\begin{figure}[htbp]
\center
\includegraphics[width=\textwidth]{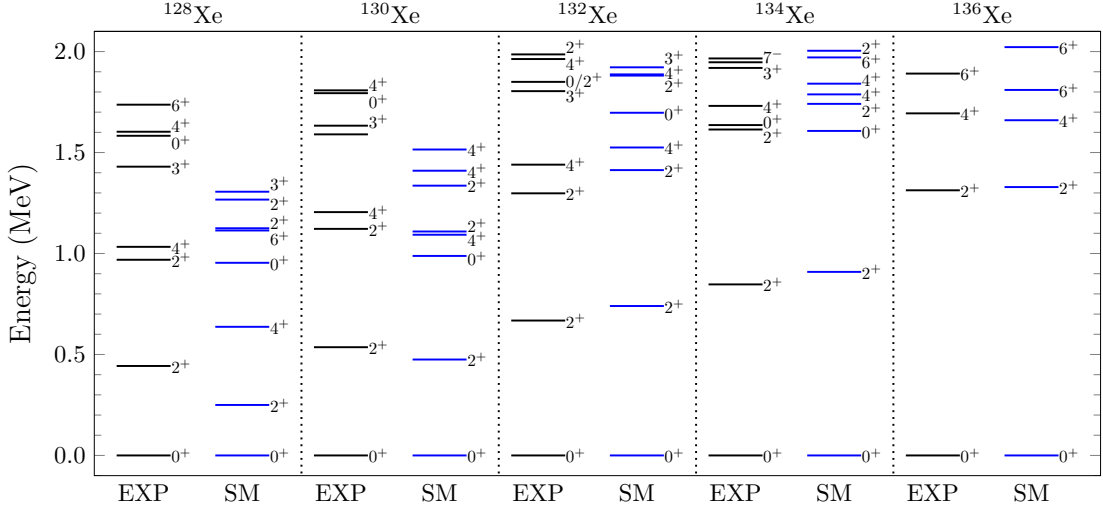}
\caption{Experimental and shell-model energy spectra of even-mass xenon isotopes. A maximum of eight lowest energy levels are shown for each isotope. From left to right: $^{128}$Xe, $^{130}$Xe, $^{132}$Xe, $^{134}$Xe, and $^{136}$Xe.}\label{fig:xeeven}
\end{figure}

\begin{figure}[htbp]
\center
\includegraphics[width=0.7\textwidth]{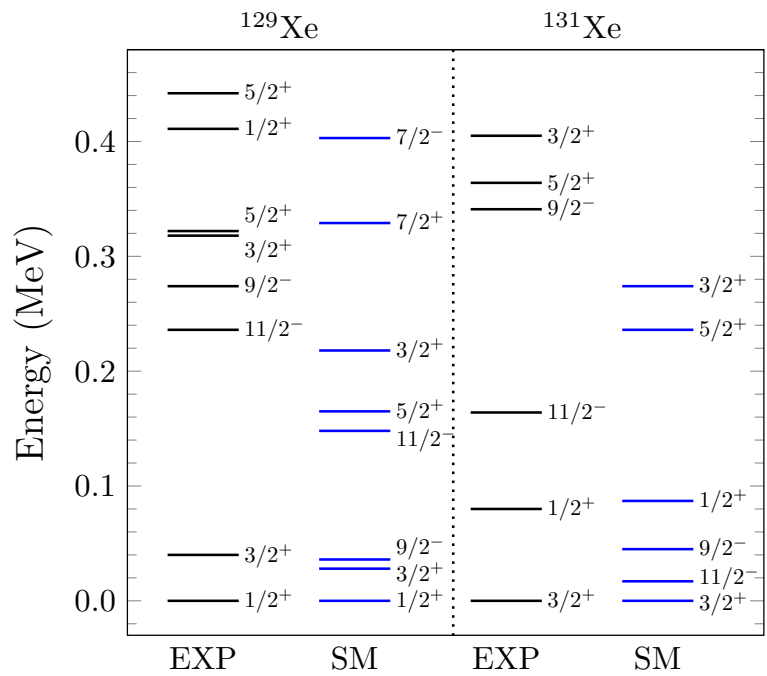}
\caption{Experimental and shell-model energy spectra of odd-mass xenon isotopes $^{129}$Xe (left) and $^{131}$Xe (right).} \label{fig:xeodd}
\end{figure}

The computed energy levels of the even-mass xenon isotopes are given in Fig. \ref{fig:xeeven} and the odd-mass isotopes in Fig. \ref{fig:xeodd}. For the even-mass isotopes the experimental energy spectra are very well reproduced by the shell-model calculations. The accuracy is somewhat diminished when moving to lower masses from the closed neutron major shell of $^{136}$Xe, but a decent correspondence between experimental and theoretical levels can be found. For the odd-mass isotopes the situation is more complex, but the positive-parity states are well reproduced by the calculations. However, the negative-parity states $11/2^-$ and $9/2^-$ are computed to be much lower than in the experimental spectrum. This effect has been observed in earlier calculations using the SN100PN interaction in this mass region \cite{Pirinen2016}. The experimental data for the xenon isotopes was obtained from \cite{nndc}.

To give a further measure of accuracy of our calculation, we computed the ground state magnetic moments for $^{129}$Xe and $^{131}$Xe. For $^{129}$Xe the experimental magnetic moment of the $1/2^+$ ground state is $\mu_\mathrm{exp}  = -0.7779763(84) \: \mu_\mathrm{N}$ while the shell-model calculated value is $\mu_\mathrm{sm} = -1.360 \: \mu_\mathrm{N}$. For $^{131}$Xe $3/2^+$ ground state the numbers are $\mu_\mathrm{exp} =+0.691862 (4) \: \mu_\mathrm{N}$ and $\mu_\mathrm{sm} = +1.059 \: \mu_\mathrm{N} $ for experiment and shell model respectively. The sign of the magnetic moment in both cases is correct, but the magnitude of both of our calculated values is somewhat larger than that of the experimental ones.

\section{Neutrino scattering results} \label{sec:results}

In this section we present the results of our calculations for neutrino-nucleus scattering cross sections by methods described in Section \ref{sec:ncscatt}. We have computed total cross sections for coherent and inelastic neutrino-nucleus scattering as a function of the neutrino energy, and also averaged total cross sections for solar $^8$B neutrinos and supernova neutrinos scattering off the most abundant xenon isotopes. In the following calculations of averaged supernova-neutrino cross sections we have used two different neutrino temperatures corresponding to different neutrino flavors. We follow the choices of Refs. \cite{Almosly2015,Almosly2013} and have the electron neutrinos described by parameters $\alpha = 3.0$, $\langle E_\nu \rangle = 11.5 \: \mathrm{MeV}$, and $T_\nu = 2.88 \: \mathrm{MeV}$, and the muon and tau neutrinos by $\alpha = 3.0$, $\langle E_\nu \rangle = 16.3 \: \mathrm{MeV}$, and $T_\nu = 4.08 \: \mathrm{MeV}$. Whenever we refer to supernova neutrinos in the following text these parameter values are used in the calculations.

\subsection{Coherent elastic scattering}

In Table \ref{tab:cohefun} we present the total cross section for coherent neutrino-nucleus scattering off the target xenon isotopes  as a function of neutrino energy. In Table \ref{tab:cohefun} we only show calculations in the nuclear shell model, but the values for the QRPA/MQPM formalism are very similar, which is reflected on the total averaged cross sections shown later. The cross sections rise rapidly for small neutrino energies and start to saturate when approaching 100 MeV. The cross sections are larger for the higher-$A$ isotopes, following the $N^2$ coherent enhancement. 

\begin{table*}[htbp]
\footnotesize
\center
\caption{Coherent elastic neutral-current scattering cross section for neutrinos  scattering off xenon targets as a function of neutrino energy. The cross sections for each isotope are given in units of $\mathrm{cm}^2/\mathrm{MeV}$ in columns 2-8 as a function of the neutrino energy (column 1). The computations were made in the nuclear shell model.} 
\label{tab:cohefun}
\begin{tabular}{llllllll} 
	\toprule \toprule
	 \multicolumn{1}{c}{$E_\nu$}                  &  \multicolumn{7}{c}{$\sigma$ (cm$^2$/MeV) }\\
          \cline{2-8} \\
	$ (\mathrm{MeV})$ & $^{128}$Xe &  $^{129}$Xe  & $^{130}$Xe & $^{131}$Xe  & $^{132}$Xe  &  $^{134}$Xe  &  $^{136}$Xe   \\
	\midrule
           5     & $5.16 \times 10^{-40}$ & $5.31 \times 10^{-40}$ & $5.46 \times 10^{-40}$ & $5.61 \times 10^{-40}$& $5.76 \times 10^{-40}$& $6.08 \times 10^{-40}$& $6.40 \times 10^{-40}$\\
           10   & $2.02 \times 10^{-39}$ & $2.08 \times 10^{-39}$ & $2.14 \times 10^{-39}$ & $2.20 \times 10^{-39}$& $2.26 \times 10^{-39}$& $2.38 \times 10^{-39}$& $2.50 \times 10^{-39}$\\
           20   & $7.44 \times 10^{-39}$ & $7.65 \times 10^{-39}$ & $7.86 \times 10^{-39}$ & $8.07 \times 10^{-39}$& $8.29 \times 10^{-39}$& $8.73 \times 10^{-39}$& $9.19 \times 10^{-39}$\\
           30   & $1.47 \times 10^{-38}$ & $1.51 \times 10^{-38}$ & $1.55 \times 10^{-38}$ & $1.59 \times 10^{-38}$& $1.63 \times 10^{-38}$& $1.71 \times 10^{-38}$& $1.80 \times 10^{-38}$\\
           40   & $2.19 \times 10^{-38}$ & $2.25 \times 10^{-38}$ & $2.31 \times 10^{-38}$ & $2.37 \times 10^{-38}$& $2.43 \times 10^{-38}$& $2.55 \times 10^{-38}$& $2.67 \times 10^{-38}$\\
           50   & $2.80 \times 10^{-38}$ & $2.88 \times 10^{-38}$ & $2.94 \times 10^{-38}$ & $3.02 \times 10^{-38}$& $3.09 \times 10^{-38}$& $3.24 \times 10^{-38}$& $3.39 \times 10^{-38}$\\
           60   & $3.25 \times 10^{-38}$ & $3.33 \times 10^{-38}$ & $3.40 \times 10^{-38}$ & $3.49 \times 10^{-38}$& $3.57 \times 10^{-38}$& $3.73 \times 10^{-38}$& $3.91 \times 10^{-38}$\\
           70   & $3.55 \times 10^{-38}$ & $3.64 \times 10^{-38}$ & $3.72 \times 10^{-38}$ & $3.81 \times 10^{-38}$& $3.89 \times 10^{-38}$& $4.07 \times 10^{-38}$& $4.25 \times 10^{-38}$\\
           80   & $3.75 \times 10^{-38}$ & $3.84 \times 10^{-38}$ & $3.92 \times 10^{-38}$ & $4.02 \times 10^{-38}$& $4.10 \times 10^{-38}$& $4.28 \times 10^{-38}$& $4.48 \times 10^{-38}$\\
\bottomrule \bottomrule
\end{tabular}
\end{table*}

We present the total averaged cross section for supernova neutrinos as well as solar $^{8}$B neutrinos in Table \ref{tab:totalcs}. Results  for coherent scattering are shown for the shell model and QRPA/MQPM calculations. The results between the shell model and quasiparticle approaches are very similar. Some small differences can be observed in the results for the odd-mass isotopes, but those are still not very significant. The cross sections for the supernova neutrinos are larger than for $^{8}$B neutrinos by roughly a factor of 3 or 5 depending on the neutrino flavor. This is due to the average energy of the supernova neutrinos being larger at 11.5 MeV or 16.3 MeV, while the $^{8}$B spectrum peaks at around 7 MeV.

\begin{table*}[htbp]
\footnotesize
\center
\caption{Total averaged cross section for $^{8}$B solar neutrinos and  electron and muon/tau supernova neutrinos (SN$\nu_\mathrm{e}$/SN$\nu_\mathrm{x}$) scattering off xenon targets. The results are shown for calculations in the nuclear shell model (SM) and the QRPA/MQPM formalisms. Cross sections for coherent scattering are given in units of $10^{-39} \: \mathrm{cm}^2$, and for inelastic scattering in $10^{-43} \: \mathrm{cm}^2$.  } \label{tab:totalcs}
\begin{tabular}{llllllll} 
	\toprule \toprule
		      &                  & $\langle \sigma \rangle_{\mathrm{coh,^{8}B}}$     &   $\langle \sigma \rangle_{\mathrm{coh,SN}\nu_\mathrm{e}}$   &  $\langle \sigma \rangle_{\mathrm{coh,SN}\nu_\mathrm{x}}$ &   $\langle \sigma \rangle_{\mathrm{inel,^{8}B}}$   &    $\langle \sigma  \rangle_{\mathrm{inel,SN}\nu_\mathrm{e}}$  &    $\langle \sigma  \rangle_{\mathrm{inel,SN}\nu_\mathrm{x}}$\\
	 Nucleus & Model     & $ (10^{-39} \: \mathrm{cm}^2)$   & $ (10^{-39} \: \mathrm{cm}^2)$  & $ (10^{-39} \: \mathrm{cm}^2)$ & $ (10^{-43} \: \mathrm{cm}^2)$    & $ (10^{-43} \: \mathrm{cm}^2)$  & $ (10^{-43} \: \mathrm{cm}^2)$      \\
	\midrule
           $^{128}$Xe & SM & 1.064  & 3.051  & 5.692 & - & -  & -\\
			 & QRPA & 1.065  & 3.052 & 5.696 & 1.567 & 38.10 & 152.0\\
           $^{129}$Xe & SM &  1.095  & 3.138 & 5.853 & - & - &-\\
			 & MQPM &  1.105  & 3.166 & 5.903  & 2.208 & 45.11 & 173.4\\
           $^{130}$Xe & SM &  1.125 & 3.223 &  6.008 & - & -&-\\
			 & QRPA & 1.126 & 3.225 & 6.013 & 1.564 & 40.94 & 161.0\\
           $^{131}$Xe & SM &  1.157 & 3.313 & 6.173 & - & - &-\\
			 & MQPM & 1.167 & 3.336 & 6.215 & 3.699 & 54.14 & 195.4\\
           $^{132}$Xe & SM &  1.188 & 3.401 & 6.335 & - & -&-\\ 
			 & QRPA & 1.189 & 3.403 & 6.339 & 2.341 & 48.21 & 180.4\\
           $^{134}$Xe & SM &  1.253 & 3.585  & 6.671 & - & -&-\\
			 & QRPA &  1.253 & 3.585 & 6.673 & 3.107 & 56.10 & 201.7\\
           $^{136}$Xe & SM &  1.320 & 3.773 & 7.016  & - & - &-\\
			 & QRPA &  1.320 & 3.773 & 7.016  & 2.102 & 53.43 & 200.5 \\
     
\bottomrule \bottomrule
\end{tabular}
\end{table*}

\subsection{Inelastic scattering}
Due to the limitations of the shell model in describing high-lying excited states, we compute inelastic scattering properties using only the QRPA/MQPM formalism, which is known to well depict the collective properties of excited nuclear states. The total cross section as a function of neutrino energy is given in Table \ref{tab:inelefun} for each xenon isotope. For smaller neutrino energies, 0 to 30 MeV, the cross sections are slightly larger for the odd-mass isotopes than for their neighboring isotopes. The energies of solar neutrinos fit completely into this range, which leads to the averaged cross sections for solar neutrinos to be larger for the odd-mass isotopes. 

\begin{table*}[htbp]
\footnotesize
\centering
\caption{Inelastic neutral-current scattering cross section for neutrinos  scattering off xenon targets as a function of neutrino energy. The cross sections are given in units of $\mathrm{cm}^2/\mathrm{MeV}$. The computations were made in the QRPA/MQPM formalism.} 
\label{tab:inelefun}
\begin{tabular}{llllllll} 
	\toprule \toprule
	 $E_\nu \: (\mathrm{MeV})$ & $^{128}$Xe    & $^{129}$Xe  & $^{130}$Xe & $^{131}$Xe  & $^{132}$Xe  &  $^{134}$Xe  &  $^{136}$Xe   \\
	\midrule
           5   & $1.71\times 10^{-45}$  & $2.10 \times 10^{-45}$  & $1.29 \times 10^{-46}$ & $2.74 \times 10^{-44}$ & $7.74 \times 10^{-45}$ & $1.28\times 10^{-44}$  & $2.00 \times 10^{-47}$ \\
           10 & $3.56 \times 10^{-43}$  & $5.27 \times 10^{-43}$  & $3.54 \times 10^{-43}$  & $8.49 \times 10^{-43}$ & $5.49 \times 10^{-43}$ & $7.57 \times 10^{-43}$ & $4.75 \times 10^{-43}$ \\
           20 & $1.41 \times 10^{-41}$  & $1.71 \times 10^{-41}$   & $1.53 \times 10^{-41} $ & $2.00 \times 10^{-41}$& $1.78 \times 10^{-41}$ & $2.06 \times 10^{-41}$ & $2.02 \times 10^{-41}$ \\
           30 & $6.50 \times 10^{-41}$  & $7.45 \times 10^{-41}$ & $6.85 \times 10^{-41}$  & $8.18 \times 10^{-41}$ & $7.53 \times 10^{-41}$ & $8.29 \times 10^{-41}$ & $8.41 \times 10^{-41}$\\
           40 & $1.85 \times 10^{-40}$  & $1.94 \times 10^{-40}$  & $1.91 \times 10^{-40}$ & $2.05 \times 10^{-40}$ & $2.02 \times 10^{-40}$ & $2.16 \times 10^{-40}$ & $2.20 \times 10^{-40}$\\
           50 & $3.99 \times 10^{-40}$  &  $3.85 \times 10^{-40}$ & $4.05 \times 10^{-40}$ & $3.95 \times 10^{-40}$ & $4.20 \times 10^{-40}$ & $4.38 \times 10^{-40}$ & $4.47 \times 10^{-40}$\\
           60 & $7.17 \times 10^{-40}$  & $6.41\times 10^{-40}$  & $7.20 \times 10^{-40}$ & $6.45 \times 10^{-40} $ & $7.35 \times 10^{-40}$ & $7.55 \times 10^{-40}$ & $7.66 \times 10^{-40}$\\
           70 & $1.14 \times 10^{-39}$  & $9.51 \times 10^{-40}$  & $1.14 \times 10^{-39}$ & $9.44 \times 10^{-40}$ & $1.15 \times 10^{-39}$ & $1.16 \times 10^{-39}$  & $1.18 \times 10^{-39}$\\
           80 & $1.66 \times 10^{-39}$  & $1.30 \times 10^{-39}$  & $1.64 \times 10^{-39}$ & $1.28 \times 10^{-39}$ & $1.65 \times 10^{-39}$ & $1.66 \times 10^{-39}$ & $1.67 \times 10^{-39}$\\     
\bottomrule \bottomrule
\end{tabular}
\end{table*}

The total averaged inelastic cross sections are listed in Table \ref{tab:totalcs}. The inelastic scattering cross sections are some orders of magnitude smaller than the coherent cross sections, as expected. Here the cross sections of the supernova neutrinos are an order of magnitude or two larger than of $^{8}$B solar neutrinos, again due to the supernova neutrinos having on average a higher energy. The effect of neutrino energy appears more pronounced in inelastic scattering than in coherent scattering, however. The cross sections of the odd-mass isotopes are again slightly larger than those of the neighboring isotopes. 

We can compare our inelastic scattering results with those calculated in Ref. \cite{Almosly2015} for Cd isotopes using the same supernova neutrino parameters. The results for Cd isotopes in Ref. \cite{Almosly2015} in the case of electron neutrino range from $4.38 \times 10^{-42} \: \mathrm{cm}^2$ for $^{106}$Cd to $4.96 \times 10^{-42} \: \mathrm{cm}^2$ for $^{111}$Cd, with a general decreasing trend with increasing mass number for even-mass nuclei. Our results for Xe isotopes in Table \ref{tab:totalcs} are very similar in magnitude, but the trend is rather rising than decreasing with increasing mass number. This could be a shell effect, as adding neutrons to Cd isotopes takes the nucleus further away from a closed major shell, but for the xenon nuclei it gets closer to a shell closure. Same conclusions can be made for the other neutrino flavors.

\begin{figure}[htbp]
\center
\includegraphics[width = \textwidth]{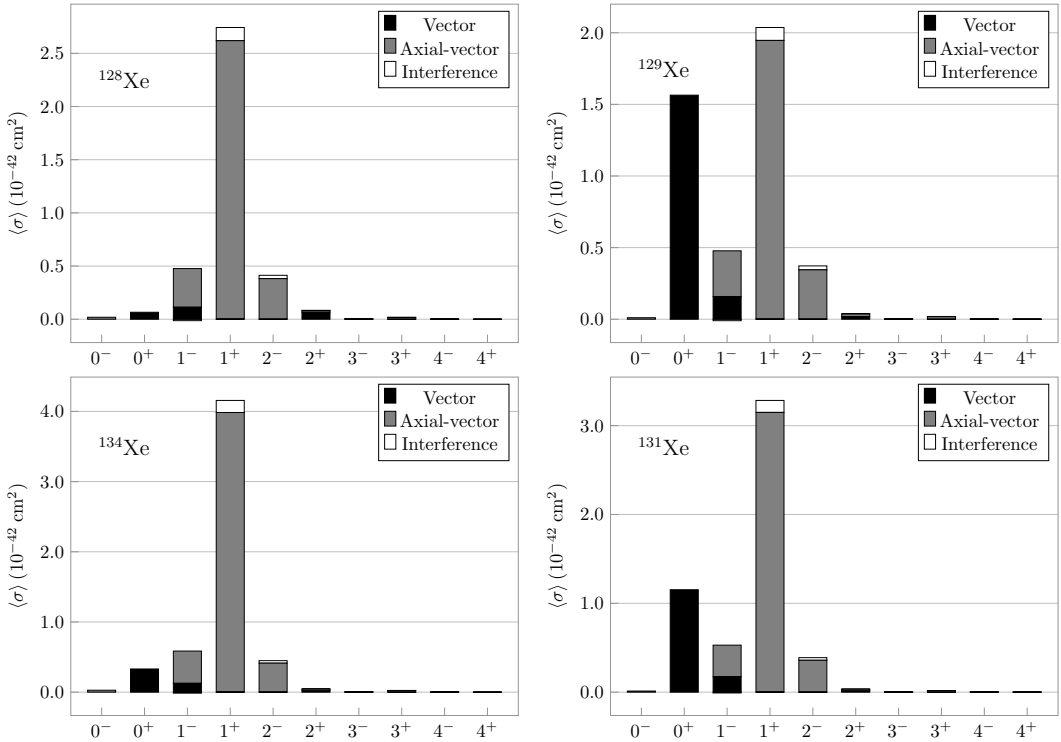}
\caption{The contributions of multipole channels $J \leq 4$ to the total averaged cross section for inelastic scattering of supernova electron neutrinos. Bar plots are shown for a representative sample of $^{128}$Xe (top left), $^{129}$Xe (top right), $^{134}$Xe (bottom left), and $^{131}$Xe (bottom right). A division to vector, axial-vector, and interference parts of the interaction is shown. Cross sections are given in units of $10^{-42} \: \mathrm{cm}^2$. } \label{fig:mult}
\end{figure}

\begin{figure}[htbp]
\center
\includegraphics[width = \textwidth]{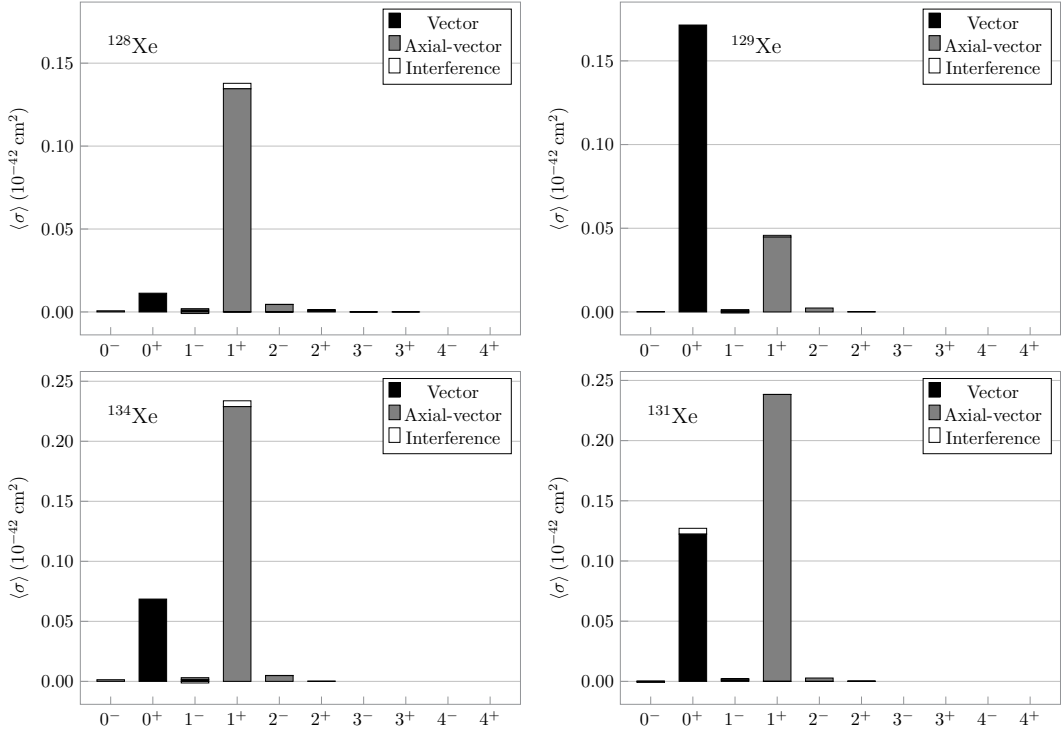}
\caption{The contributions of multipole channels $J \leq 4$ to the total averaged cross section for inelastic scattering of solar $^{8}$B neutrinos. Bar plots are shown for a representative sample of $^{128}$Xe (top left), $^{129}$Xe (top right), $^{134}$Xe (bottom left), and $^{131}$Xe (bottom right). A division to vector, axial-vector, and interference parts of the interaction is shown. Cross sections are given in units of $10^{-42} \: \mathrm{cm}^2$.   } \label{fig:multsolar}
\end{figure}

We show the contributions from different multipole channels to the total averaged cross sections in Fig. \ref{fig:mult} for supernova electron neutrinos and Fig. \ref{fig:multsolar} for solar neutrinos. It is evident that the most dominant contribution comes from an axial-vector $1^+$ multipole transition in all cases but one. Smaller, yet still important contributions arise from the axial-vector $1^-$ and $2^{-}$ channels for higher neutrino energies. This is characteristic behavior for neutral-current scattering, which has been observed in Ref. \cite{Almosly2015} for Cd isotopes, and in Ref. \cite{Ydrefors2012} for Mo isotopes. The contributions get more evenly distributed among the different multipoles with increasing neutrino energy.  

For the odd-mass nuclei our calculations also show a significant contribution from a vector $0^{+}$ channel, and for solar neutrinos scattering off $^{129}$Xe this channel in fact becomes the strongest. For the even-mass isotopes this channel is more suppressed, but it becomes more significant for the lower energy solar neutrinos. Similar large $0^+$ contributions were observed in \cite{Almosly2015} for Cd isotopes. This is problematic as, in principle, the $0^+$ contribution is expected to be small because it vanishes at the limit $q \to 0$. The particle-number violation of the quasiparticle framework can be an explanation for the large computed $0^+$ contribution. A detailed examination on the origins of the $0^+$ anomaly will be conducted in a later study. At this time one should regard the $0^+$ contributions with caution as they are probably at least partially spurious.

\begin{figure}[htbp]
\center
\includegraphics[width=\textwidth]{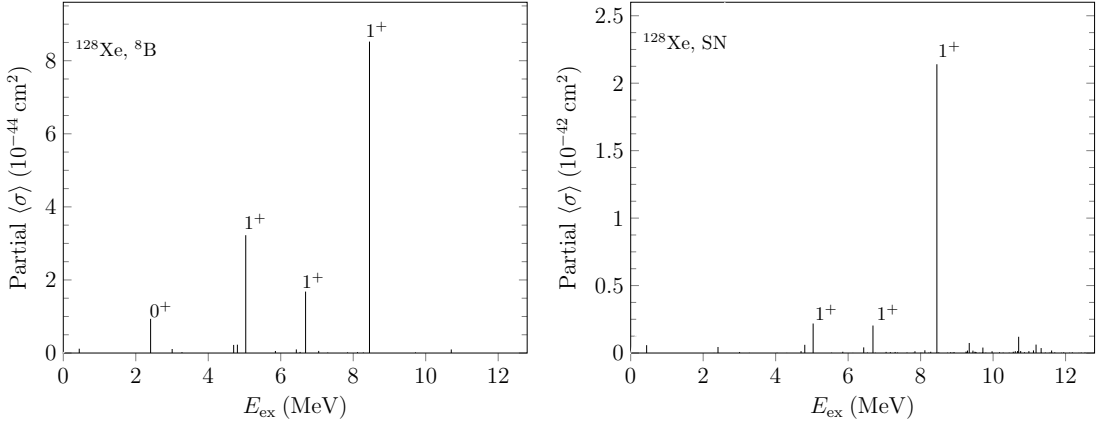}
\caption{Contributions to the inelastic scattering averaged cross section arising from various final states of $^{128}$Xe at energies $E_\mathrm{ex}$. Results are shown for $^{8}$B solar neutrinos (left panel) and supernova electron neutrinos (right panel). Cross sections are given in units of $10^{-44} \: \mathrm{cm}^2$ for solar neutrinos, and $10^{-42} \: \mathrm{cm}^2$ for supernova neutrinos.} \label{fig:omega128}
\end{figure}

\begin{figure}[htbp]
\center
\includegraphics[width=\textwidth]{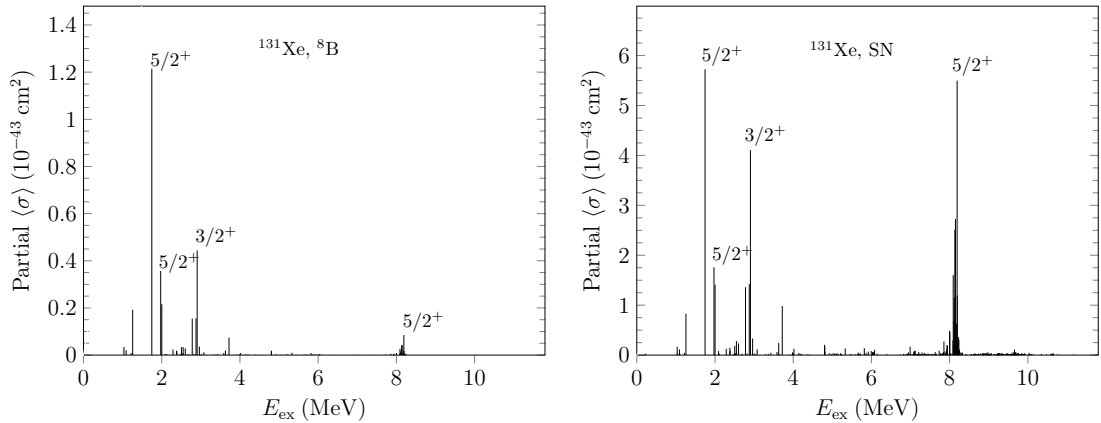}
\caption{Contributions to the inelastic scattering averaged cross section arising from various final states of $^{131}$Xe at energies $E_\mathrm{ex}$. Results are shown for $^{8}$B solar neutrinos (left panel) and supernova electron neutrinos (right panel). In both panels cross sections are given in units of $10^{-43} \: \mathrm{cm}^2$.}
\label{fig:omega131}
\end{figure}

In Figs. \ref{fig:omega128} and \ref{fig:omega131} we show the dominating contributions to the inelastic scattering cross section from various final states of $^{128}$Xe and $^{131}$Xe, respectively. We notice that the major contributions are very similar for the solar and supernova electron neutrinos for the even-mass $^{128}$Xe, where the leading contributions come from $1^{+}$ states at  8.4 MeV, 5.0 MeV, and 6.7 MeV. For solar neutrinos there is also a notable contribution from a $0^+$ state at 2.4 MeV. The situation is very much different for the odd-mass $^{131}$Xe, where for supernova neutrinos there is a pile of $5/2^+$, $3/2^+$, and $1/2^+$ states at roughly 8 MeV giving large contributions to the total cross section in addition to the large contributions from lower-lying $5/2^+$ and $3/2^+$ states. However, for solar $^{8}$B neutrinos this peak at 8 MeV is much smaller, and the leading contributions are more localized to the $5/2^+$ state at 1.8 MeV and the $3/2^+$ state at 2.9 MeV. It is interesting that a relatively small change in the average neutrino energy can lead to the higher-lying states to give much larger contributions to the total cross section.

\begin{figure}[htbp]
\center
\includegraphics[width=\textwidth]{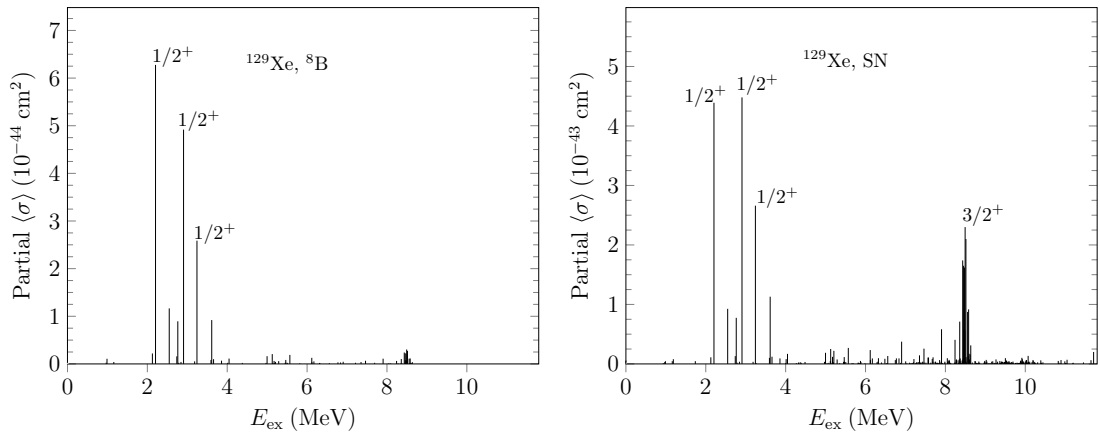}
\caption{Contributions to the inelastic scattering averaged cross section arising from various final states of $^{129}$Xe at energies $E_\mathrm{ex}$. Results are shown for $^{8}$B solar neutrinos (left panel) and supernova electron neutrinos (right panel). Cross sections are given in units of $10^{-44} \: \mathrm{cm}^2$ for solar neutrinos, and $10^{-43} \: \mathrm{cm}^2$ for supernova neutrinos.} \label{fig:omega129}
\end{figure}

Following the discussion on the anomalously large $0^+$ multipole contribution in $^{129}$Xe we show the dominant final states for neutrinos scattering off $^{129}$Xe in Fig. \ref{fig:omega129}. As expected from the large $0^+$ multipole, the largest contributions here come from $1/2^+$ states at energies of roughly $2-3$ MeV. Something in the nuclear-structure calculation seems to favor the $0^+$ multipole transition to $1/2^+$ final states over the $1^+$ multipole transition to $3/2^+$ states. Otherwise similar conclusions can be made for $^{129}$Xe as for $^{131}$Xe above about the location of the peaks in energy and differences between solar and supernova neutrinos.

\section{Conclusions} \label{sec:conc}

We have computed various properties of cross sections of neutral-current neutrino-nucleus scattering off the most abundant Xe isotopes. The nuclear structure of our target Xe nuclei was computed in the nuclear shell model for elastic scattering, and in the QRPA framework for both elastic and inelastic scattering. For the odd-mass nuclei $^{129}$Xe and $^{131}$Xe an MQPM calculation was performed based on the QRPA calculation for $^{130}$Xe and $^{132}$Xe, respectively. We used realistic neutrino energy distributions for solar neutrinos from $^{8}$B beta decay and supernova neutrinos to compute the averaged cross sections for each neutrino scenario.

The total averaged cross section for supernova neutrinos are dependent of the values of the parameters $\alpha_\nu$ and $\langle E_\nu \rangle$. We have shown results of only one set of parameters for electron neutrinos and one for muon/tau neutrinos. The dependence of the cross sections on the parameter $\alpha_\nu$ is typically quite mild, unless the change is large \cite{Ydrefors2015,Almosly2015}. The values $\alpha_\nu = 3.0$, $\langle E_{\nu_\mathrm{e}} \rangle = 11.5 \: \mathrm{MeV}$, and $\langle E_{\nu_\mathrm{x}} \rangle = 16.3 \: \mathrm{MeV}$ used in this work are reasonable estimates and allow comparison of results with the $^{8}$B solar neutrinos, for which the energy distribution is better known. A mapping of cross sections for various supernova neutrino parameters is out of scope of this work. However, we have tabulated total cross section as a function of neutrino energy, which can be used to obtain estimates of total averaged cross sections for any neutrino energy profile.

The scattering process in even-even nuclei is dominated by transitions to high-lying $1^+$ states, and for odd-mass nuclei typically by states differing from the initial state by one unit of angular momentum. We found that in even-mass nuclei the leading contributions from various final states are quite similar between solar neutrinos and supernova neutrinos. In odd-mass nuclei, however, the smaller energy of the solar neutrinos does not allow large contributions to the total cross section to arise from high-lying states. We also noted that the smaller energies of solar neutrinos lead into an enhancement in the vector $0^+$ multipole channel in comparison to the otherwise dominating $1^+$ axial-vector channel, especially in the odd-mass Xe nuclei. However, the large contribution from the $0^+$ multipole can be mostly spurious, possibly due to the particle number violation of the quasiparticle framework. This matter will be investigated further and subsequently reported elsewhere.

\section*{Acknowledgments}\label{thanks}

This work has been partially supported by the Academy of Finland under 
the Finnish Centre of Excellence Programme 2012-2017 (Nuclear and 
Accelerator Based Programme at JYFL). P. Pirinen 
was supported by a graduate student stipend from the Magnus Ehrnrooth Foundation. 
E. Ydrefors thanks for the financial support of the grant \#2016/25143-7 from the Sao Paulo Research Foundation (FAPESP).

\end{document}